\documentstyle [aps,prc,preprint]{revtex}
\preprint{KFA-IKP(TH)-1996-09}
\begin {document}
\draft

\title
{Meson-meson scattering:\\ $K\overline{K}$-thresholds and 
$f_{0}(980) - a_{0}(980)$ mixing}

\author 
{O. Krehl$^{1}$, R. Rapp$^{2}$, and J. Speth$^{3,1}$}

\address
{1)  Institut f\"{u}r Kernphysik, Forschungszentrum
     J\"{u}lich GmbH, \\ D-52425 J\"{u}lich, Germany \\
 2)  Department of Physics, State University of New York at Stony Brook,\\
     Stony Brook, New York 11794, USA \\
 3)  Thomas Jefferson National Accelerator Facility \\
     12000 Jefferson Ave., Newport News, VA 23606, USA}

\maketitle

\begin{abstract}
We study the influence of mass splitting between the 
charged and neutral pions and kaons in the J\"ulich meson 
exchange model for $\pi\pi$ and $\pi\eta$ scattering. The 
calculations are performed in the particle basis, which permits 
the use of physical masses for the pseudoscalar mesons and a 
study of the distinct thresholds associated with the neutral and the 
charged kaons. Within this model we also investigate the isospin violation 
which arises from the mass splitting and an apparent violation 
of $G$-parity in $\pi\pi$ scattering which stems from the 
coupling to the $K\overline{K}$ channel. Nonvanishing cross sections for 
$\pi\pi \rightarrow \pi^{0}\eta$ indicate a mixing of the $f_{0}(980)$ and 
$a_{0}(980)$ states.
\end{abstract}
\pacs{11.80.Gw, 13.75.Lb, 14.40.Cs, 13.40.Dk}

\section{introduction}
In strong interaction physics isospin is a nearly exact 
symmetry weakly broken by the slightly 
different masses within isospin multiplets - for example between proton 
and neutron, or between charged and neutral pions or kaons. For that reason 
it is in general a good 
approximation to describe baryon-baryon, meson-baryon and 
meson-meson scattering in an isospin basis with explicit account taken
of Coulomb effects. In phase shift analyses of scattering processes isospin 
is taken to be a conserved quantum number.

The J\"ulich model for meson-meson interactions~\cite{janssen,lohse} is 
constructed in the same spirit, {\it i.e.} the mass splitting between 
neutral and charged pseudoscalar mesons is neglected and the scattering 
equation is solved in the isospin basis. The corresponding interaction 
kernel is based on an effective Lagrangian which includes the scalar, 
pseudoscalar and vector nonets. An important feature of the model is the 
coupling between $\pi\pi$ and $K\overline{K}$ which in the scalar-isoscalar 
channel ($JI$=00) gives rise to a bound $K\overline{K}$ state 
($K\overline{K}$-molecule~\cite{weinstein}). This 'dynamical pole' is the 
$f_{0}(980)$ 'meson', which therefore strongly decays into the 
$K\overline{K}$ channel. 
The decay into $K\overline{K}$ is under experimental investigation at 
the cooler-synchrotron COSY \cite{cosy}. Because of the high momentum 
resolution of COSY the thresholds from neutral 
and charged kaons are well separated and can be studied independently.

An appropriate theoretical description of the $K^0\overline{K^0}$/$K^+K^-$ 
threshold region has to account for their kinematical separation. In this 
article we therefore reformulate the J\"ulich 
$\pi\pi$/$K\overline{K}$ model in a basis specified by mass eigenstates 
which permits inclusion of the mass splitting between charged and 
neutral pseudoscalar mesons. These mass differences give rise to a violation 
of isospin. Furthermore, we also include the $\pi\eta$ channel 
in our calculations. It's coupling to the $K\overline{K}$ channel is crucial 
for the understanding of the $a_{0}(980)$ meson~\cite{janssen,weinstein}.
As it will turn out, the such constructed $\pi\pi - K\overline{K} - \pi\eta$
model not only accounts for isospin symmetry breaking effects but also leads
to a mixing of states with different $G$-parity.  
This surprising result has a very simple physical origin: in the particle 
basis the $\pi\pi$ and $\pi\eta$ channels couple through the $K\overline{K}$ 
channel which does not have a definite G-parity. Intrinsically this mixing is 
very small. However, as we will see, the effect is  enhanced in the region 
of the $K\overline{K}$-molecule by the resonance structure; 
this gives rise to quite a strong mixing of the $\pi\pi$ ($\pi^{0}\pi^{0}$ 
and $\pi^{+}\pi^{-}$) and $\pi\eta$ channel.

In the next section we briefly review the J\"ulich meson-meson model and 
give the pertinent formulas for the extension to the particle basis. The 
results of our calculations are presented in sect.~\ref{secresults} 
together with a discussion of the related physics issues. In 
sect.~\ref{secsummary} we summarize and address future 
applications and extensions.

\section{The J\"ulich model of meson-meson interactions}\label{secmodel}
Our starting point is the J\"ulich $\pi\pi$ and $\pi\eta$ 
model~\cite{janssen,lohse}, which is constructed from an effective meson 
lagrangian. The corresponding potentials include t-/u-channel $\rho$ 
exchange for $\pi\pi \rightarrow \pi\pi$, $K^*$(892) exchange for 
$\pi\pi, \pi\eta \rightarrow K\bar{K}$ as well as $\rho$, $\omega$ and $\phi$ 
exchange for $K\bar{K} \rightarrow K\bar{K}$. Furthermore, s-channel 
polegraphs for $\rho$, $f_2$(1270) and $\epsilon$(1200-1400)  
formation are necessary to reproduce the resonance structures in the 
corresponding partial waves. The scalar-isoscalar particle $\epsilon$ is an 
effective description of the singlet and octet member of the scalar nonet 
as well as a possible glueball. \\  
The $\pi\pi\rho$ coupling constant $g_{\pi\pi\rho}$ for t-/u-channel 
$\rho$ exchange is determined by the decay width of the $\rho$-meson. 
On the other hand, s-channel resonances 
are renormalized by solving the scattering eq.~(\ref{scateq}) (see below),
so that the corresponding bare coupling constants and masses have to be 
determined by reproducing experimental information. All other coupling 
constants are given from SU(3) symmetry relations. For the regularization of 
the scattering eq.~(\ref{scateq}) the vertex functions are supplemented with 
hadronic form factors which we choose of dipole form:  
\begin{eqnarray}
F(\vec k,\vec k') &=& \left(\frac{2\Lambda^{2}-M^2}{2\Lambda^2+(\vec k - 
\vec k')^{2}}\right)^{2}
\quad \mbox{for t-,u-channel} \nonumber \\
F(k) &=& \left(\frac{2\Lambda^{2}+M^2}{2\Lambda^2+4\omega_k^2}\right)^{2} 
\quad \mbox{for s-channel} \nonumber \ ,      
\end{eqnarray}
where the cutoffs $\Lambda$ are adjusted to experimental data. 
The parameters of the model \cite{janssen} are given in 
tables~\ref{parameter}, \ref{nackt}.

The scattering amplitudes are calculated by iterating the potentials in a 
coupled channel scattering equation which we obtain from the 
Blankenbecler-Sugar (BbS) reduction~\cite{bbs} of the Bethe-Salpeter 
equation. After partial wave expansion, one has for given isospin $I$ and 
angular momentum $J$:   
\begin{equation}\label{scateq}
T_{\mu\nu}^{IJ}(k,k',E) = V_{\mu\nu}^{IJ}(k,k',E) + \sum_\lambda
\int dk''k''^2 V_{\mu\lambda}^{IJ}(k,k'',E) G_\lambda(k'',E) 
T_{\lambda\nu}^{IJ} (k'',k',E) \ ,
\end{equation}
where $\mu,\nu,\lambda  = \pi\pi, K\bar{K}$ for the $\pi\pi$ model 
($\mu,\nu,\lambda = \pi\eta,K\bar{K}$ for the $\pi\eta$ model) and  
$E=\sqrt{s}$ is the starting energy.
The BbS two-meson propagator for channel $\lambda$ reads  
\begin{equation}
G_\lambda(k'',E) = \frac{\omega_1+\omega_2}{(2\pi)^3 2\omega_1\omega_2} \ 
\frac{1}{E^2 - (\omega_1+\omega_2)^2 + i\eta} \ ,
\end{equation}
with $\omega_{1/2} = \sqrt{k''^2 + m_{1/2}^2}$. The on-shell momentum 
$k_0$ is defined as
\begin{equation}
k_0 = \frac{\sqrt{[E^2-(m_1+m_2)^2][E^2-(m_1-m_2)^2]}}{2E}.
\end{equation}

Let us now turn to the transformation from isospin basis to particle basis.  
The physical particles are represented by  
\begin{eqnarray*}
\pi^+\pi^- &=& -|11\rangle |1-1\rangle \\
\pi^0\pi^0\; &=& \quad |10\rangle |10\rangle \\
K^+K^- &=& -|\frac{1}{2} \frac{1}{2} \rangle |\frac{1}{2} -\frac{1}{2} \rangle
\end{eqnarray*}
and so on. The phase convention is taken from the SU(3) representation of the 
pseudoscalar octet, where $|\pi^+\rangle = -|11\rangle$ and 
$|K^-\rangle = -|\frac{1}{2}-\frac{1}{2}\rangle$; all other particles are 
identified with a plus sign. The potentials are then transformed according to
\begin{equation}
\langle \pi^+\pi^-|V^J| \pi^+\pi^- \rangle = \sum_{I=0}^{2} \langle 
\pi^+\pi^-| I I_z \rangle
\langle I I_z |V^{IJ}| I I_z \rangle \langle I I_z | \pi^+\pi^- \rangle \ ,
\end{equation}
where $\langle \pi^+\pi^-| I I_z \rangle$ are Glebsch-Gordan coefficients. 
A factor of $\sqrt{2}$ must be multiplied to matrix elements transforming 
isospin states of identical particles (such as $|\pi\pi\rangle$) to 
distinguishable particle states (e.g. $|\pi^+ \pi^- \rangle$) in order to  
compensate the factor of $\frac{1}{\sqrt{2}}$ arising from the normalization 
of states of identical particles, {\it e.g.} 
\begin{eqnarray}\label{trafo}
\langle \pi^+\pi^-|V^J| \pi^+\pi^- \rangle &=& \frac{2}{3} 
V^{I=0,J}_{\pi\pi\rightarrow\pi\pi} + 
V^{I=1,J}_{\pi\pi \rightarrow\pi\pi} + 
\frac{1}{3}V^{I=2,J}_{\pi\pi\rightarrow\pi\pi} \\ \nonumber
\langle \pi^+\pi^-|V^J| \pi^0\pi^0 \rangle &=& \frac{\sqrt{2}}{3} 
V^{I=0,J}_{\pi\pi\rightarrow\pi\pi} - 
\frac{\sqrt{2}}{3}V^{I=2,J}_{\pi\pi\rightarrow\pi\pi}
\end{eqnarray}
and so on. As a consequence of Bose symmetry, the u-channel graphs cancel 
the corresponding t-channel graphs for $(-1)^{I+J}=-1$, but give the same 
contribution if $(-1)^{I+J}=1$.  
So we include u-channel graphs by multipying t-channel graphs with 
a factor of 2 and neglecting the whole process if $(-1)^{I+J}=-1$.  
{\it E.g.} for S-waves, $V^{I=1,J=0}_{\pi\pi\rightarrow\pi\pi}$ does 
not contribute. 
The transformation is completed by employing the physical masses  
in the coupled channel formalism. \\    
Since isospin is no longer conserved we additionally include $\pi\eta$ states 
in the scattering equation. Restricting ourselves to charge neutral 
initial and final states leads to five channels   
$\mu,\nu,\lambda  = \pi^0\pi^0,\pi^+\pi^-,\pi^0\eta,K^+K^-$ and 
$K^0\bar{K^0}$ in eq.~(\ref{scateq}).

To determine the poles of the scattering amplitude $T_{\mu\nu}$ in the 
complex energy plain we follow the approach of 
Janssen et. al.~\cite{janssen}. Due to the mass splitting of pions and 
kaons and the inclusion of $\pi\eta$ states we now have 5 right hand 
cuts with distinct thresholds. Thus, we extend the sheet notation using a 
5 character string composed of the 
two letters $b$ and $t$ to denote whether the energy lies on the bottom 
or top sheet of the corresponding cut. The cuts are ordered with 
increasing energy, so the first, second, third, fourth and fifth 
character corresponds to the $\pi^{+}\pi^{-}$, $\pi^{0}\pi^{0}$, 
$\pi^{0}\eta$, $K^{+}K^{-}$ and $K^{0}\overline{K^{0}}$ channel, 
respectively.

\section{Numerical results and physics issues}\label{secresults}
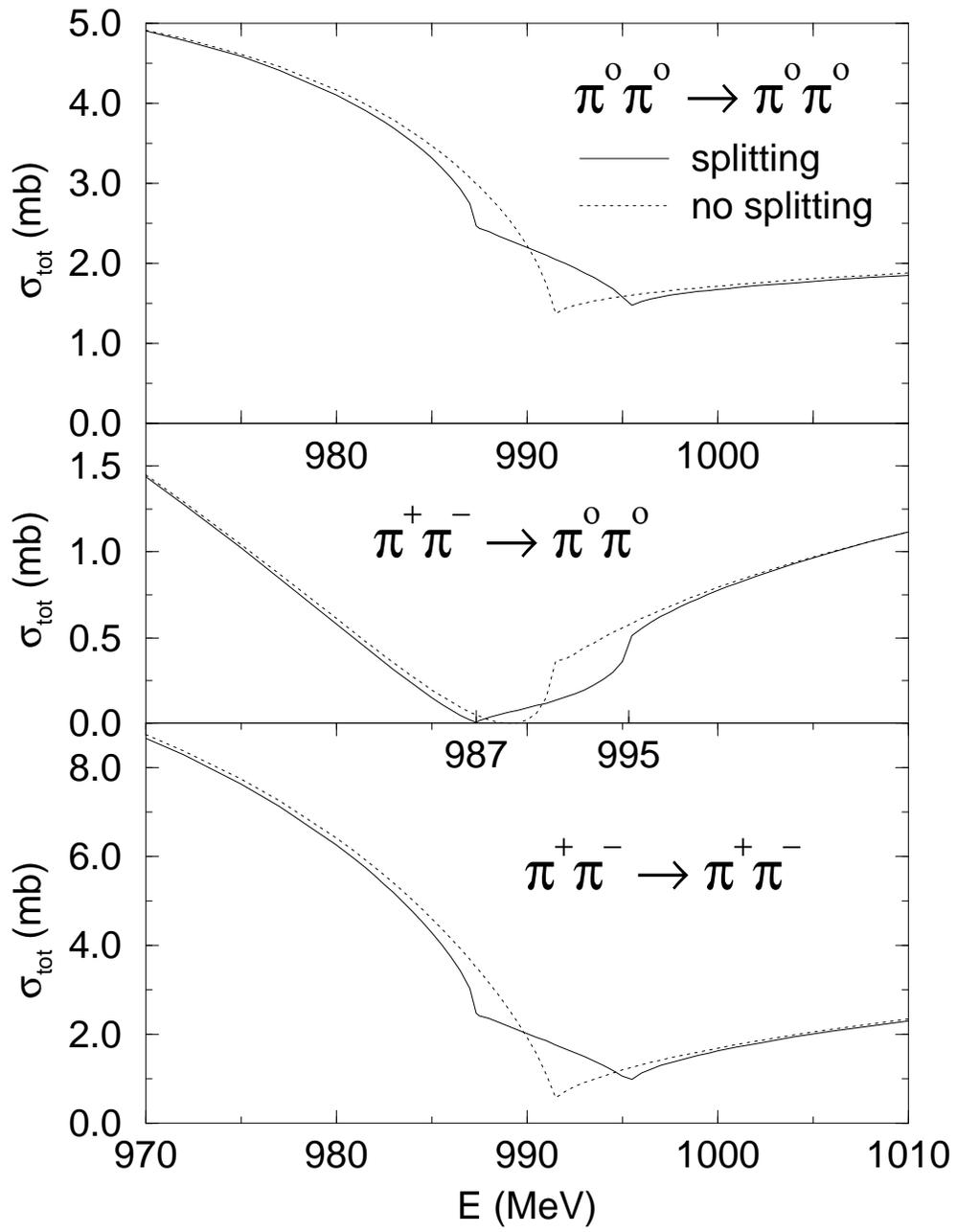
\begin{figure}
\caption{S-wave cross sections for $\pi\pi$ scattering. The solid 
lines are calculated using physical masses, whereas for the dotted 
lines average masses ($m_{\pi^{0}}= m_{\pi^{+}}=137.273$ MeV, 
$m_{K^{+}}= m_{K^{0}}=495.675$ MeV) are employed.}
\label{fig1}
\end{figure}

The numerical calculations are performed in a similar way as reported 
in ref.~\cite{janssen}. The main difference is the larger number of coupled 
channels which have to be considered. 
We concentrate in our present investigation on the $K\overline{K}$ threshold 
region around 1~GeV where the scalar mesons $f_{0}(980)$ and $a_{0}(980)$ are 
located. Here the influence of the pion mass differences is negligible small. 
All the effects we are discussing here stem from the mass splitting between 
the neutral and charged kaons.
In fig.~\ref{fig1} we show the various $\pi\pi$ scattering cross sections 
for $J$=0 in the charge-neutral channels in the vicinity of the 
$K\overline{K}$ threshold. All three cross sections exhibit the well 
known interference pattern related to the $f_{0}(980)$ meson which, 
however, in the present model is a $K\overline{K}$ molecule and not a 
genuine $q\overline{q}$ state. The full and 
dashed lines represent the results obtained 
with and without mass splitting between the charged and neutral kaons. 
If the mass splitting is considered we obtain two kinks in the cross 
sections rather than one. These (predicted) discontinuities arise from 
the $K^{+}K^{-}$ and $K^{0}\overline{K^{0}}$ thresholds and will be 
investigated with the cooler synchrotron COSY in the near future. 
This prominent two-kink structure is directly connected with a strong 
interaction in the $K\overline{K}$ channel which gives rise to a 
$K\overline{K}$ molecule. Therefore an experimental verification of this 
structure will strongly support the $K\overline{K}$ molecule 
character of the $f_{0}(980)$.

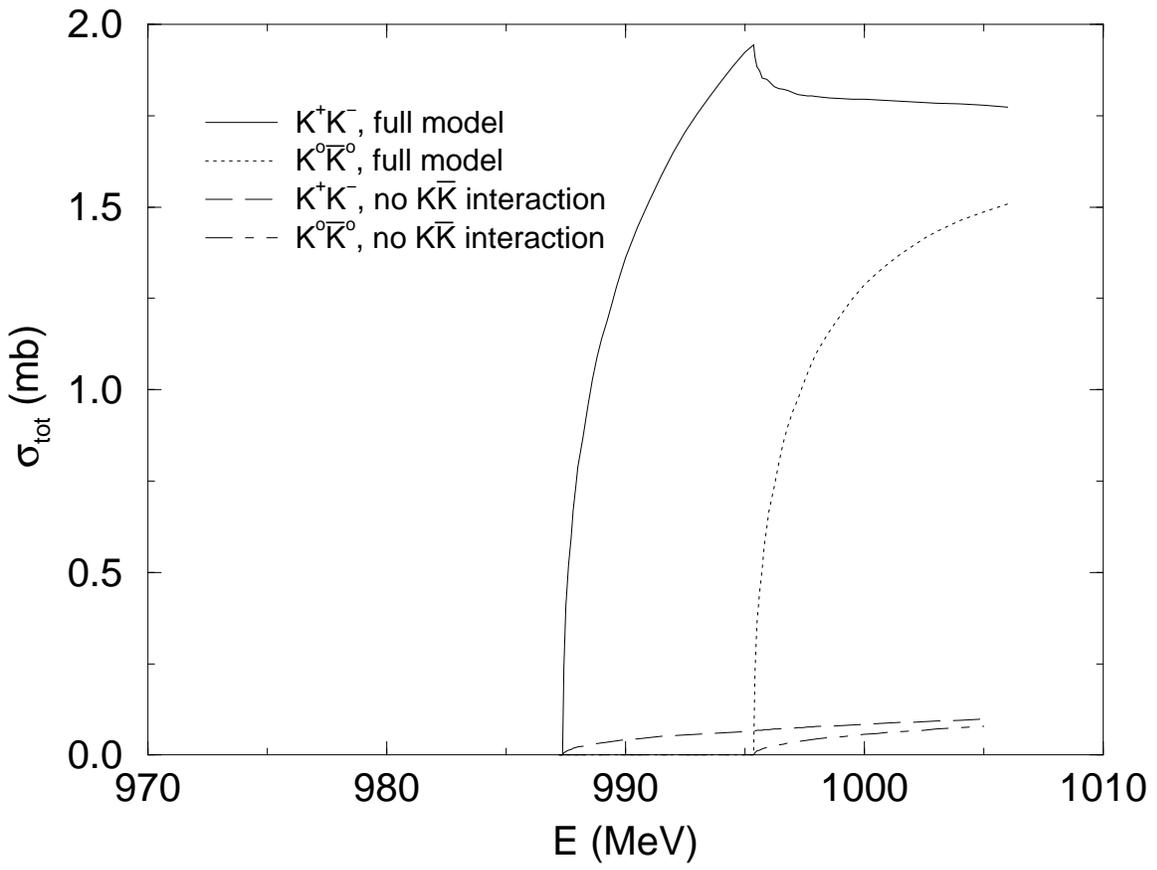
\begin{figure}
\caption{S-wave cross sections for $\pi\pi \rightarrow K\overline{K}$. 
The solid/dotted line corresponds to $\pi^{+}\pi^{-} \rightarrow 
K^{+}K^{-}/K^{0}\overline{K^{0}}$, respectively. The dashed/dotted-dashed line shows
the cross section neglecting the $K\overline{K}$ interaction. The cross sections for
$\pi^{0}\pi^{0} \rightarrow K\overline{K}$ can be obtained by multiplying $\sigma_
{\pi^{+}\pi^{-} \rightarrow K\overline{K}}$ with $\frac{1}{2}$.}
\label{fig2}
\end{figure}

In fig.~\ref{fig2} we show the threshold behavior of the  
cross sections for the transitions $\pi\pi \rightarrow 
K^{0}\overline{K^{0}}, K^{+}K^{-}$. The full and dotted lines are the 
results of the full model whereas the dashed and dotted-dashed lines arise 
from calculations where the $K\overline{K}$ interaction is neglected. 
The energy dependence of these cross sections 
near threshold is again directly linked to the $K\overline{K}$ interaction. 
Therefore the analyses of experiments as {\it e.g.} presently performed at 
COSY~ \cite{cosy} give a quite model-independent measure of the 
$K\overline{K}$ interaction and, correspondingly, insight into the structure 
of the $f_{0}(980)$ meson.

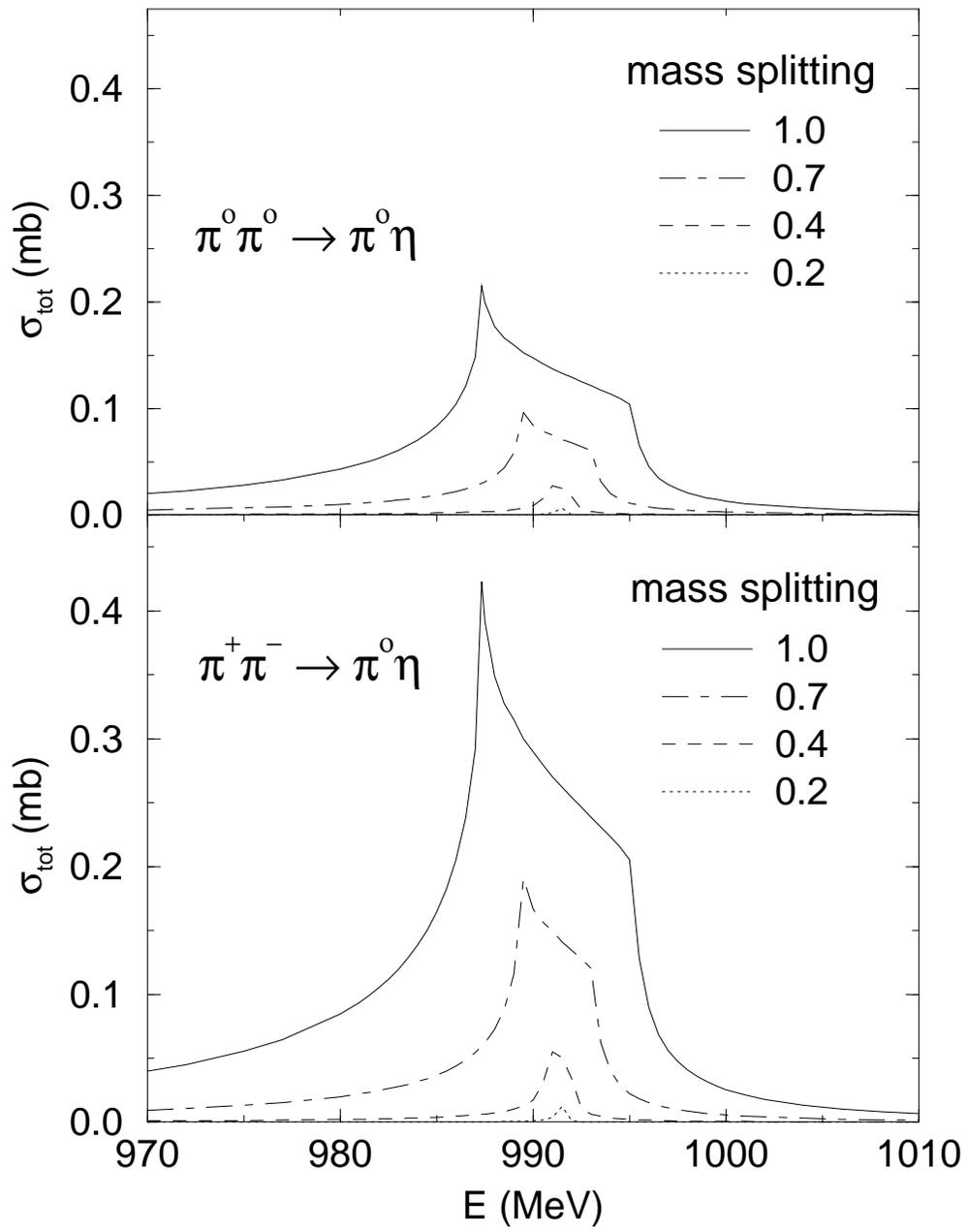
\begin{figure}
\caption{Our prediction for the transition cross sections 
$\pi\pi \rightarrow \pi^0\eta$ for $J=0$}
\label{fig3}
\end{figure}

A quite unexpected result is shown in fig.~\ref{fig3} where 
the transition cross sections $\pi\pi \rightarrow \pi\eta$ are displayed. 
Such a transition is only possible if not only isospin but also $G$-parity is 
no longer a conserved quantum number. The isospin violation comes 
from the mass splitting of the charged and neutral kaons. 
The apparent violation of $G$-parity in $\pi\pi$ scattering actually 
stems from the coupling to the $K\overline{K}$ channel. Whereas pions 
posses $G=-1$, kaons do not have a definite $G$-parity because 
$K^{+}, K^{0}$ and $K^{-}, \overline{K^{0}}$ belong to two different 
isospin doublets (the $G$-parity transformation is defined as a 
rotation of 180 degrees about the y-axis in isospin space followed by 
charge conjugation). From fig.~\ref{fig3} we clearly see that the 
$G$-parity mixing (violation) is again a phenomenon which is directly 
connected with the resonance structure at the $K\overline{K}$ threshold. 
Only in the vicinity of the threshold the transition cross section is 
relatively large but it rapidly ceases  when moving away from the 
$K\overline{K}$ resonance. If we neglect the $K\overline{K}$ 
interaction this resonance enhancement does not show up. In 
fig.~\ref{fig4} we show the dependence of the transition cross 
section on the magnitude of the kaon mass splitting
which is a measure for the isospin violation. 
As one can see, the effect depends very strongly 
on the magnitude of the mass splitting. As mentioned before, the isospin 
violation due to the mass splitting of the pions does not have any impact  
on this energy region.

The origin of the $K\overline{K}$ molecule can be seen most clearly by 
determining the pole positions of the scattering amplitude $T_{\mu\nu}$ 
in the complex energy plane. Here a bound state pole is a single pole 
on a sheet near the physical region, which moves to the real axis
as the coupling to the other channels decreases. 
At zero coupling the bound state
pole is located on the top sheet below threshold on the real axis. 
In our model we find two single poles at energies $[bbbtt] 
(1014.1,\pm14.1)$~MeV and $[bbttt] (1014.2,\pm13.9)$~MeV.
When neglecting the $\pi^{0}\eta$ channel these two poles move to the same 
energy $(1014.2,\pm13.8)$~MeV on both sheets $[bbbtt]$ and $[bbttt]$, 
{\it i.e.} the $\pi^{0}\eta$ channel has only a very minor effect on the 
$f_{0}(980)$ bound state leaving two (almost) degenerate poles. Next we 
remove the $\epsilon$ s-channel contribution: the two poles are mildly 
affected moving to $[bbtt] (1020.0,\pm59.4)$~MeV (no $\pi^{0}\eta$ 
channel included). The last step is the decoupling of the $\pi\pi$ 
channel, which can be achieved by continously decreasing the 
$\pi\pi \rightarrow K\overline{K}$ coupling.
By doing so, the pole moves below threshold onto the real axis and ends up 
on all $\pi\pi$ sheets at $[xytt] (986.8,\pm0.0)$~MeV ($x,y=b,t$) indicating 
a $K\overline{K}$ molecule. 

This pole analysis in particle basis fully confirmes the findings in 
isospin basis. The same is true for the $a_{0}(980)$ where we also 
find that it originates from the $\pi\eta \rightarrow K\overline{K}$
transition (cusp effect, see ref.~\cite{janssen}). 

Our theoretical finding that the $K\overline{K}$ molecule can decay into 
$\pi\pi$ as well as into $\pi\eta$ sheds new light on the structure of the 
$a_{0}(980)$ meson. In all previous analyses the $\pi\eta$ decay was 
considered as an unambiguous signal for the $a_{0}$ meson. In the light 
of our present result this assumption seems to be no longer justified. 
It can well be that at least in some of the experiments the $\pi^{0}\eta$ 
decay may have had it's origin from the $f_{0}$(980)-'meson'. Thus, the 
fact that in some experiments the width of the $a_{0}$ and $f_{0}$ 
appeared to be very similar could be naturally explained by our findings 
and would mean that in both cases the width of the $K\overline{K}$ 
molecule has been measured. In this respect it is 
interesting to note that in the calculations of ref.~\cite{janssen} only the  
$f_{0}(980)$ turns out to be a bound $K\overline{K}$ molecule whereas 
the $a_{0}(980)$ does not:  even though the latter is strongly affected by 
the $K\overline{K}$ threshold, the t-channel vector meson exchanges in the 
$I$=1 are not attractive enough to create a bound state. Therefore the 
enhancement of the $K\overline{K}$ state in the isovector channel ($I$=1) 
is much less pronounced than in the isoscalar one ($I$=0).
This means that the $\pi^{0}\eta$ decay of the $f_{0}(980)$ can simulate 
an $a_{0}$(980) 'meson', but the $\pi\pi$ decay of the $a_{0}(980)$ will 
hardly influence the appearance of $f_{0}(980)$. 

\section{summary}\label{secsummary}
In the present article we have studied the influence of the mass splitting 
between charged and neutral pions and kaons in $\pi\pi$, $K\overline{K}$ and 
$\pi\eta$ scattering. For this purpose we have generalized the J\"ulich meson 
exchange model for meson-meson scattering: the calculations have been 
performed in particle basis rather than in isospin basis which permits 
the use of physical masses for the pseudoscalar mesons. Our interest was 
focused on the distinct thresholds associated with the neutral and charged 
kaons and on the $f_{0}(980)$ and $a_{0}(980)$ mesons which, at least in 
our model, are directly related to them. The two thresholds can be 
clearly separated in the $J$=0 $\pi\pi$ cross sections where they give 
rise to the well known interference pattern around 1~GeV. The structure 
of the corresponding pole, which is responsible for this interference 
and which one identifies with the $f_{0}(980)$ meson, is essentially 
unchanged compared to the calculations in isospin basis of 
ref.~\cite{janssen}: it remains a $K\overline{K}$ bound state (molecule). 
We have also calculated the transition cross sections 
$\pi\pi \rightarrow K^{+}K^{-}$ and 
$\pi\pi \rightarrow K^{0}\overline{K^{0}}$. The slope of these cross sections 
depends sensitively on the direct $K\overline{K}$ interaction. 
The analysis of the forthcoming data~\cite{cosy} will give important 
information on the magnitude of this interaction and,  
connected with it, on the structure of the scalar mesons in this region.

Within our model we have furthermore investigated the isospin violation which 
arises from the mass splitting and, most interestingly, an apparent violation 
of $G$-parity in $\pi\pi$ scattering which stems from the coupling to the 
$K\overline{K}$ channel. The simultaneous violation of isospin and 
$G$-parity gives rise to nonvanishing cross 
sections for $\pi\pi \rightarrow \pi\eta$ indicating a mixing 
of $f_{0}(980)$ and $a_{0}(980)$ states. This has important consequences 
for the analysis and interpretation of 
the $a_{0}(980)$ data and it may well be that most of the $\pi^{0}\eta$ decay 
observed in this region originates from the $f_{0}(980)$ rather than from 
the $a_{0}(980)$ meson. In this context we are currently investigating  
pion production in high-energy $\pi^{-}p$ reactions where the $f_{0}(980)$ 
and $a_{0}(980)$ can be seen. The results 
will be published elsewhere.

\acknowledgements

It is a pleasure to thank Alex Dzierba for several discussions on the 
experimental situation which stimulated some of the present investigations. 
Two of us (O.K. and J.S.) wish to thank Nathan Isgur for discussions and the 
hospitality they enjoyed during their visits at Jefferson Lab. This work was 
supported in part by the US Department of Energy under contract 
DE-AC05-84ER40150

\pagebreak

\begin{figure}
\caption{The $\pi\pi \rightarrow \pi^0\eta$ cross sections when 
decreasing $K\overline{K}$ interaction. We achieve the reduction 
of the $K\overline{K}$ interaction by multiplying the
$K\overline{K} \rightarrow K\overline{K}$ potentials with a factor 
given in the legend.}
\label{fig4}
\end{figure}

\begin{table}
\caption{Vertex parameters for t- and s-channel (index (0)) graphs.}
\label{parameter}
\begin{tabular}{ccc}
Vertex & g & $\Lambda$[MeV] \\
\tableline
$\pi\pi\rho$ & 6.04 & 1355 \\
$\pi KK^*$ & $-\frac{1}{2}g_{\pi\pi\rho}$ & 1900 \\
$KK\rho$ & $\frac{1}{2}g_{\pi\pi\rho}$ & 1850 \\
$KK\omega$ & $\frac{1}{2}g_{\pi\pi\rho}$ & 2800 \\
$KK\phi$ & $\frac{1}{\sqrt{2}}g_{\pi\pi\rho}$ & 2800 \\
$\eta KK^*$ & $-\frac{\sqrt{3}}{2}g_{\pi\pi\rho}$ & 3290 \\
\tableline
$\pi\pi\epsilon^{(0)}$ & 0.286 & 1850 \\
$KK\epsilon^{(0)}$ & -0.286 & 2500 \\
$\pi\pi\rho^{(0)}$ & 5.32 & 3300 \\
$KK\rho^{(0)}$ & $\frac{1}{2}g_{\pi\pi\rho^{(0)}}$ & 2000 \\
$\pi\pi f_2^{(0)}$ & 1.002 & 2320 \\
$KKf_2^{(0)}$ & $\frac{2}{3}g_{\pi\pi f_2^{(0)}}$ & 2800 \\
\end{tabular}
\end{table}

\begin{table}
\caption{Bare masses $m_0$ used in s-channel graphs (in MeV)}
\label{nackt}
\begin{tabular}{ccc}
$\epsilon^{(0)}$ & $\rho^{(0)}$ & $f_{2}^{(0)}$ \\
\tableline
1520 & 1125 & 1665 \\
\end{tabular}
\end{table}

\end{document}